\documentclass[prl,showpacs,preprintnumbers,amsmath,amssymb,twocolumn]{revtex4}

\usepackage{graphicx}
\usepackage{dcolumn}
\usepackage{bm}

\begin{document}

\title{Doping of Mn$_2$VAl and Mn$_2$VSi Heusler alloys as a route to half-metallic antiferromagnetism}

\author{I. Galanakis$^1$}\email{galanakis@upatras.gr} \author{K.
\"Ozdo\~gan$^2$}\email{kozdogan@gyte.edu}
\author{E. \c Sa\c s\i o\~glu$^{3,4}$}\email{e.sasioglu@fz-juelich.de}
\author{B. Akta\c s$^2$}

\affiliation{$^1$ Department of Materials Science, School
of Natural Sciences, University of Patras,  GR-26504 Patra, Greece\\
$^2$ Department of Physics, Gebze Institute of Technology, Gebze,
41400, Kocaeli, Turkey\\
$^3$ Institut f\"ur Festk\"orperforschung, Forschungszentrum
J\"ulich, D-52425 J\"ulich, Germany\\
$^4$ Fatih University, Physics Department, 34500, B\" uy\" uk\c
cekmece,  \.{I}stanbul, Turkey}

\date{\today}

\begin{abstract}
Half-metallic antiferromagnets  are the ideal materials for
spintronic applications since their zero magnetization leads to
lower stray fields and thus tiny energy losses. Starting from the
Mn$_2$VAl and Mn$_2$VSi alloys we substitute Co and Fe for Mn and
we show by means of first-principle electronic structure
calculations that the resulting compounds are ferrimagnets. When
the total number of valence electrons reaches the magic number of
24 the Fe-doped compounds are semi-metals and thus non-magnetic
while the Co-doped ones show the desirable half-metallic
antiferromagnetic character. The compounds are very likely to be
synthesized experimentally since the parent compounds, Mn$_2$VAl
and Co$_2$VAl, have been already grown in the Heusler $L2_1$
lattice structure.
\end{abstract}


\maketitle

Half-metallic ferromagnets (HMF) are at the center of scientific
research during the last decade due to their potential
applications in  spintronic devices \cite{Zutic}. These materials
are ferromagnets where the one of the two spin-channels presents a
gap at the Fermi level \cite{deGroot}. Several materials have been
predicted theoretically based on first-principles calculations to
present this peculiar behavior : several Heusler alloys
\cite{Galanakis-Review}, some magnetic oxides and colossal
magnetoresistance materials \cite{Soulen}, diluted magnetic
semiconductors \cite{McDonald}, transition-metal pnictides and
chalcogenides \cite{Mavropoulos-Review}, and Heusler
semiconductors doped with high-valent transition metal atoms
\cite{Nanda}. Heusler alloys are particularly attractive for
applications due to their very high Curie temperatures and their
structural similarity to the widely used binary semiconductors
like GaAs, InP, etc.

Although the research on HMF is intense, the ideal case would be a
half-metallic antiferromagnet (HMA), also known as
fully-compensated ferrimagnet, like the hypothetical Heusler
MnCrSb \cite{Leuken} or Mn$_3$Ga \cite{Felser2006} compounds,
since such a compound would not give rise to stray flux and thus
would lead to smaller energy consumption in devices. Unfortunately
these alloys do not crystallize in the desired structure. In the
absence of HMA a lot of studies have been focused on the
half-metallic ferrimagnets (HMFi) which yield lower total spin
moments than HMF. Van Leuken and de Groot have shown that doping
of the semiconductor FeVSb results in such a material
\cite{Leuken}. Also some other perfect Heusler compounds like
FeMnSb \cite{deGroot2} and Mn$_2$VAl
 \cite{Kemal,Gala-Full,Sasioglu}  are predicted to be HMFi.
Recently other routes to half-metallic ferrimagnetism have been
studied like the doping of diluted magnetic semiconductors
\cite{Akai} and the inclusion of defects in Cr pnictides
\cite{Galanakis-RC}.

In this letter we will study another route leading to the
desirable HMA, the doping with Co of the Mn$_2$VAl and Mn$_2$VSi
which are well known to be HMFi. The importance of this route
stems from the existence of Mn$_2$VAl in the Heusler $L2_1$ phase
as shown by several groups \cite{itoh}. Each Mn atom has a spin
moment of around -1.5 $\mu_B$ and V atom a moment of about 0.9
$\mu_B$ \cite{itoh}. All theoretical studies on Mn$_2$VAl agree on
the half-metallic character with a gap at the spin-up band instead
of the spin-down band as for the other half-metallic Heusler
alloys \cite{Kemal,Gala-Full,Sasioglu,Weht}. \c Sa\c s\i o\~glu
and collaborators studied in detail the exchange interactions in
the Mn$_2$VZ (Z=Al,Ge) HMFi and showed that the antiferromagnetic
coupling between the V  and Mn atoms stabilizes the ferromagnetic
alignment of the Mn spin moments \cite{Sasioglu}. Except
Mn$_2$VAl, also the case of compounds with Ga, In, Si, Ge and Sn
instead of Al have been predicted to be HMFi \cite{Kemal}.

In the following we will use  the full--potential nonorthogonal
local--orbital minimum--basis band structure scheme (FPLO)
\cite{koepernik} in conjunction with the local density
approximation to study the properties of the
[Mn$_{1-x}$X$_x$]$_2$VAl and [Mn$_{1-x}$X$_x$]$_2$VSi compounds
where X is Co or Fe. The coherent potential approximation is
employed to ensure random doping of the lattice sites. We will
show that doping with either Fe or Co keeps the HMFi of the ideal
alloys respecting the Slater-Pauling (SP) rule (the total spin
moment in the unit cell is  the number of valence electrons minus
24 \cite{Gala-Full})\footnote{The number of valence electrons is
calculated as $2\times [(1-x)*z^{Mn}+x*z^{Co(Fe)}]+z^V+z^{Al(Si)}$
where $z$ is the number of valence electrons of the corresponding
chemical element.}. When the concentration $x$ is such that there
are exactly 24 valence electrons the Co-doped compounds show the
desirable HMA character contrary to the Fe-doped ones which loose
their magnetic character and are simple semi-metals. The Co-doped
compounds are very likely to be synthesized experimentally since
the parent compounds Mn$_2$VAl and Co$_2$VAl \cite{Carbonari}
already exist.

\begin{figure}
\includegraphics[scale=0.4]{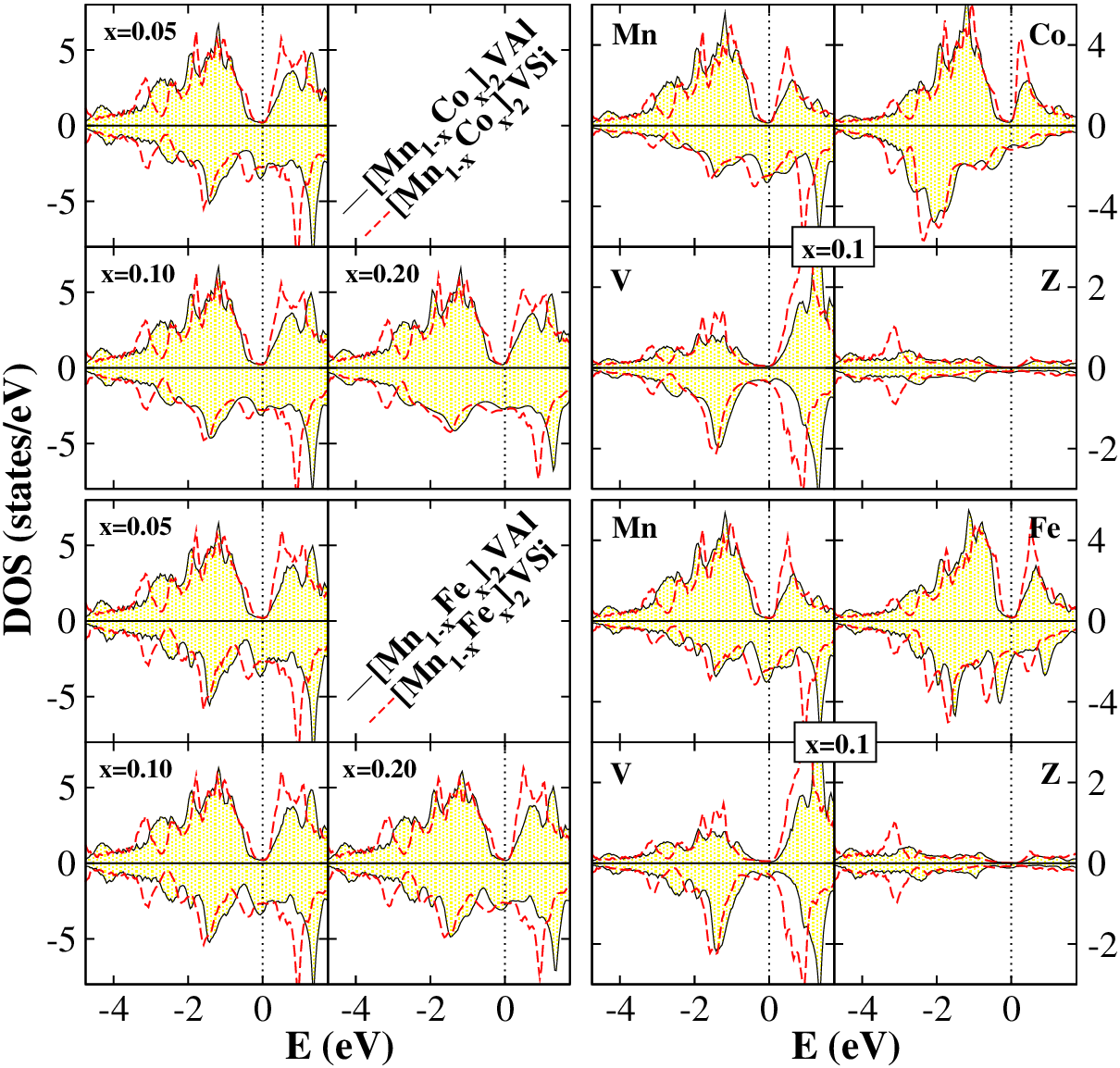}
\caption{(Color online) Top panel : total DOS as a function of the
concentration $x$ (left column) and atom-resolved DOS for $x=0.1$
(right panel) for the [Mn$_{1-x}$Co$_{x}$]$_2$VAl and
[Mn$_{1-x}$Co$_{x}$]$_2$VSi compounds. Note that the atomic DOS's
have been scaled to one atom and $Z$ corresponds either to Al or
Si. The Fermi level has been chosen as the zero of the energy
axis, and positive values of DOS correspond to the spin-up
(minority) electrons while negative values correspond to the spin-down (majority) electrons.\\
Bottom Panel : similar to the top panel for the
[Mn$_{1-x}$Fe$_{x}$]$_2$VAl and [Mn$_{1-x}$Fe$_{x}$]$_2$VSi
compounds.  \label{fig1}}
\end{figure}

\begin{figure}
\includegraphics[scale=0.4]{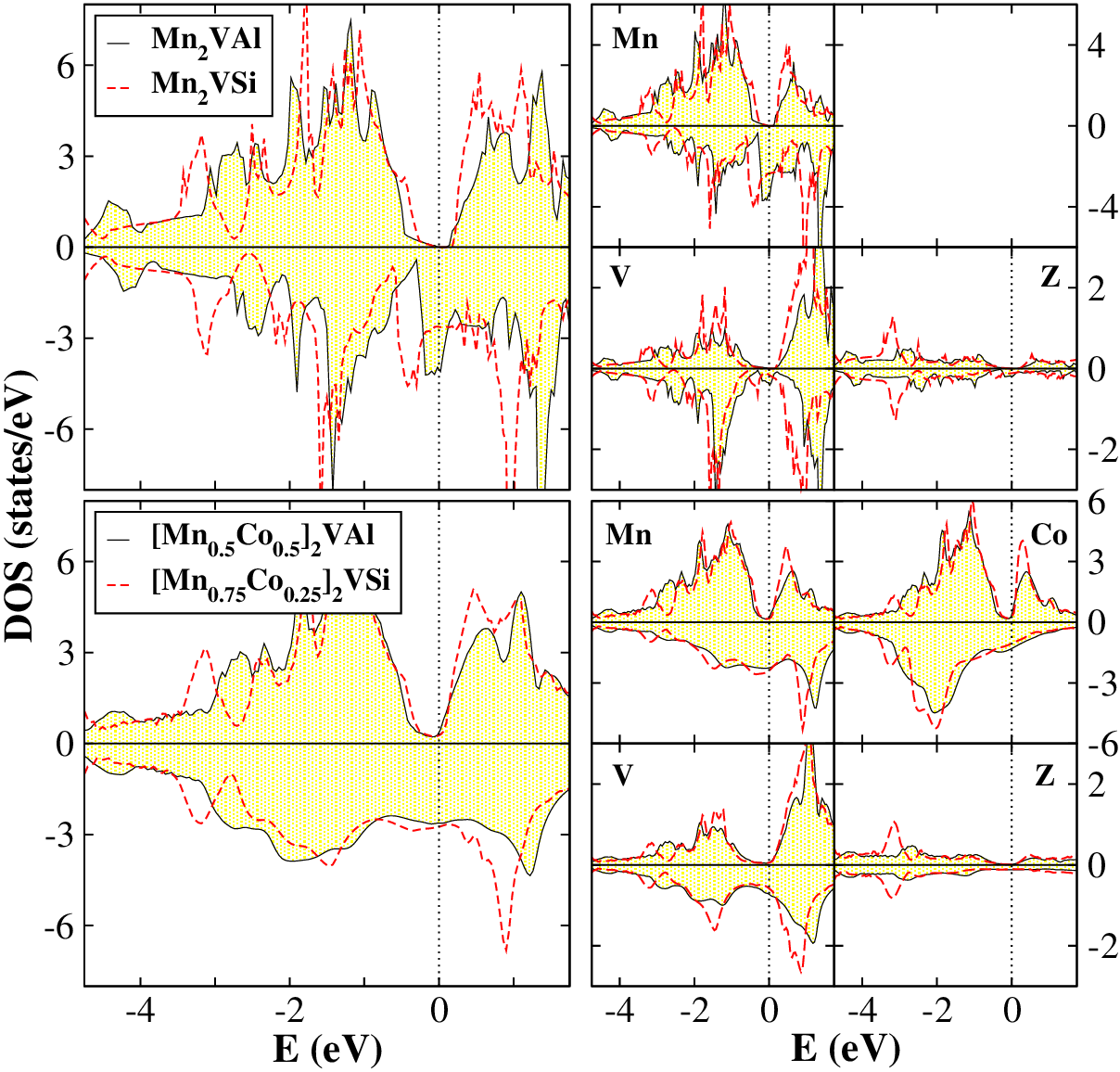}
\caption{(Color online) Top panel : total DOS (left column) and
atom resolved DOS (right panel) for the perfect Mn$_2$VAl and
Mn$_2$VSi compounds.\\
Bottom panel : similar to the top panel for  the Co-based
half-metallic antiferromagnetic  [Mn$_{0.5}$Co$_{0.5}$]$_2$VAl and
[Mn$_{0.75}$Co$_{0.25}$]$_2$VSi compounds.   \label{fig2}}
\end{figure}

We will start our discussion from the Co-doping. Prior to the
presentation of our results we have to note that due to the SP
rule \cite{Gala-Full}, these compounds with less than 24 valence
electrons have negative total spin moments and the gap is located
at the spin-up band. Moreover the spin-up electrons correspond to
the minority-spin electrons and the spin-down electrons to the
majority electrons contrary to the other Heusler alloys
\cite{Gala-Full}. We have substituted Co for Mn in Mn$_2$V(Al or
Si) in a random way and in the upper panel of Fig. \ref{fig1} we
present the total density of states (DOS) as a function of the
concentration $x$ in [Mn$_{1-x}$Co$_x$]$_2$VAl (solid black line)
and [Mn$_{1-x}$Co$_x$]$_2$VSi (red dashed line) for $x$=0.05, 0.1
and 0.2 (left column) and the atom-resolved DOS for x=0.1 in the
top right panel. The perfect compounds show a region of low
spin-up DOS (we will call it a ``pseudogap'') instead of a real
gap. Upon doping the pseudogap at the spin-up band persists and
the quaternary alloys keep the half-metallic character of the
perfect Mn$_2$VAl and Mn$_2$VSi compounds.  Co atoms are strongly
polarized by the Mn atoms since they occupy the same sublattice
and they form Co-Mn hybrids which afterwards interact with the V
and Al or Si states \cite{Gala-Full}. The spin-up Co states form a
common band with the Mn ones and the spin-up DOS for both atoms
has similar shape. Mn atoms have less weight in the spin-down band
since they accommodate less charge than the heavier Co atoms.

In Table \ref{table1} we have gathered the total and atom-resolved
spin moments for all the Co-doped compounds as a function of the
concentration. We have gone up to a concentration which
corresponds to 24 valence electrons in the unit cell, thus up to
$x$=0.5 for the [Mn$_{1-x}$Co$_x$]$_2$VAl and x=0.25 for the
[Mn$_{1-x}$Co$_x$]$_2$VSi alloys. In the last column we have
included the total spin moment predicted by the Slater-Pauling
(SP) rule for the perfect half-metals \cite{Gala-Full}. A
comparison between the calculated and ideal total spin moments
reveals that all the compounds under study are half-metals with
very small deviations due to the existence of a pseudogap instead
of a real gap. Exactly for 24 valence electrons the total spin
moment vanishes as we will discuss in the next paragraph. Co atoms
have a spin moment parallel to the V one and antiparallel to the
Mn moment, and thus the compounds retain their ferrimagnetic
character. As we increase the concentration of the Co atoms in the
alloys, each Co  has more Co atoms as neighbors, it hybridizes
stronger with them and its spin moment increases while the spin
moment of the Mn atom decreases (these changes are not too
drastic). The sp atoms have a spin moment antiparallel to the Mn
atoms as already discussed in Ref. \onlinecite{Kemal}.

\begin{table}
\caption{Atom-resolved spin magnetic moments for the
[Mn$_{1-x}$Co$_x$]$_2$VAl and [Mn$_{1-x}$Co$_x$]$_2$VSi compounds
(moments have been scaled to one atom). The two last columns are
the total spin moment (Total) in the unit cell calculated as
$2\times [(1-x)*m^{Mn}+x*m^{Co}]+m^V+m^{Al\:or\:Si}$ and the ideal
total spin moment predicted by the SP rule for half-metals (see
Ref. \onlinecite{Gala-Full}). The lattice constants have been
chosen 0.605 nm for Mn$_2$VAl and 0.6175 for Mn$_2$VSi for which
both systems are half-metals (see Ref. \onlinecite{Kemal}) and
have been kept constant upon Co doping. } \label{table1}
\begin{ruledtabular}
 \begin{tabular}{lcccccc}
 \multicolumn{7}{c}{[Mn$_{1-x}$Co$_x$]$_2$VAl} \\
 $x$ & Mn   & Co & V & Al & Total &Ideal \\
  0    & -1.573  & -- & 1.082  & 0.064 & -2.000 & -2.0 \\

0.025 & -1.587 & 0.406 & 1.102 & 0.074 & -1.899 & -1.9 \\

0.05 & -1.580 & 0.403 & 1.090 & 0.073 & -1.799 & -1.8 \\

0.1 & -1.564 & 0.398 & 1.067 & 0.069 & -1.600 & -1.6 \\

0.2 & -1.522 & 0.412 & 1.012 & 0.059 & -1.200 & -1.2 \\

0.3 & -1.484 & 0.456 & 0.953 & 0.047 & -0.804 & -0.8 \\

0.4 & -1.445 & 0.520 & 0.880 & 0.034 & -0.404 & -0.4 \\

0.5& -1.388 & 0.586 & 0.782 & 0.019 & $\sim$0 & 0\\

\multicolumn{7}{c}{[Mn$_{1-x}$Co$_x$]$_2$VSi} \\
 $x$ & Mn   & Co & V & Si & Total &Ideal \\

0 &  -0.960 & -- & 0.856 & 0.063&-1.000 &  -1.0 \\

0.025 & -0.958 & 0.716 & 0.870 & 0.062 & -0.900 & -0.9 \\

0.05 & -0.944 & 0.749 & 0.860 & 0.059 & -0.800 & -0.8 \\

0.1 & -0.925 & 0.819 & 0.847 & 0.054 & -0.600 & -0.6 \\

0.2 & -0.905& 0.907 & 0.839 & 0.046 & -0.201 & -0.2 \\

0.25&-0.899 & 0.935 & 0.839 & 0.041 & $\sim$0 & 0

\end{tabular}
\end{ruledtabular}
\end{table}

The most interesting point in this substitution procedure is
revealed when  we increase the Co concentration to a value
corresponding to 24 valence electrons in the unit cell, thus  the
[Mn$_{0.5}$Co$_{0.5}$]$_2$VAl and [Mn$_{0.75}$Co$_{0.25}$]$_2$VSi
alloys. SP rule predicts for these compounds a zero total spin
moment in the unit cell and  the electrons population is equally
divided between the two spin-bands. Our first-principles
calculations reveal that this is actually the case. The interest
arises from the fact that although the total moment is zero, these
two compounds are made up from  strongly magnetic components. Mn
atoms have a mean spin moment of $\sim$-1.4 $\mu_B$ in
[Mn$_{0.5}$Co$_{0.5}$]$_2$VAl and $\sim$-0.9 $\mu_B$ in
[Mn$_{0.75}$Co$_{0.25}$]$_2$VSi. Co and V have spin moments
antiferromagnetically coupled to the Mn ones which for
[Mn$_{0.5}$Co$_{0.5}$]$_2$VAl are  $\sim$0.6 and $\sim$0.8
$\mu_B$, respectively, and for [Mn$_{0.75}$Co$_{0.25}$]$_2$VSi
$\sim$0.9 and $\sim$0.8 $\mu_B$. Thus these two compounds are
half-metallic fully-compensated ferrimagnets or as they are best
known in literature half-metallic antiferromagnets. To confirm the
HMA character of these two compounds in Fig. \ref{fig2} we have
drawn the total and atom resolved DOS of both compounds (bottom
panel) together with the DOS of the parent Mn$_2$VAl and Mn$_2$VSi
compounds (upper panel). The substitution of 50\% of the Mn atoms
by Co ones in  Mn$_2$VAl leads to a smoothening of both the total
and atom-projected DOS due to the hybridization between the Mn and
Co atoms. Overall the energy position of the Mn states does not
change and the Fermi level falls within the pseudogap in the
spin-up band and in a region of high-DOS in the spin-down band.
All the remarks drawn in the previous paragraphs are still valid.
A similar picture occurs also when substituting 25\% of the Mn
atoms by Co in Mn$_2$VSi. The case of
[Mn$_{0.5}$Co$_{0.5}$]$_2$VAl is of particular interest since both
Mn$_2$VAl \cite{itoh} and Co$_2$VAl \cite{Carbonari} exist
experimentally in the $L2_1$ structure of Heusler alloys and this
quaternary compounds seems very likely to be synthesized.

\begin{table}
\caption{Same as table 1 for the Fe doping of the Mn-sites. Note
that when the total number of valence electrons is 24 we get a non
magnetic semi-metal.} \label{table2}
\begin{ruledtabular}
 \begin{tabular}{lcccccc}
 \multicolumn{7}{c}{[Mn$_{1-x}$Fe$_x$]$_2$VAl} \\
 $x$ & Mn   & Fe & V & Al & Total &Ideal \\
   0    & -1.573  & -- & 1.082  & 0.064 & -2.000 & -2.0 \\



0.1 & -1.604 & -0.179 & 1.054 & 0.071 & -1.799 & -1.8 \\

0.2 & -1.602 & -0.222 & 0.987 & 0.066 & -1.599 & -1.6 \\


0.4 & -1.572 & -0.242 & 0.827 & 0.054 & -1.199 & -1.2\\


0.6 &-1.498 & -0.210 & 0.616 & 0.039 & -0.796 & -0.8 \\

0.8 &-1.315 & -0.136 & 0.337 & 0.020 & -0.387 & -0.4 \\

1.0 & \multicolumn{6}{c}{non-magnetic semi-metal} \\

\multicolumn{7}{c}{[Mn$_{1-x}$Fe$_x$]$_2$VSi} \\
 $x$ & Mn   & Fe & V & Al & Total &Ideal \\
0 &  -0.960 & -- & 0.856 & 0.063&-1.000 &  -1.0 \\



0.1 & -0.979 & 0.487 & 0.808 & 0.055 & -0.800 & -0.8 \\

0.2 & -0.961 & 0.437 & 0.718 & 0.045 & -0.600 & -0.6 \\

0.3 & -0.899 & 0.367 & 0.605 & 0.034 & -0.400 & -0.4 \\

0.4 & -0.726 & 0.257 & 0.446 & 0.021 & -0.200 & -0.2 \\

0.5 & \multicolumn{6}{c}{non-magnetic semi-metal}

\end{tabular}
\end{ruledtabular}
\end{table}

\begin{figure}
\includegraphics[scale=0.5]{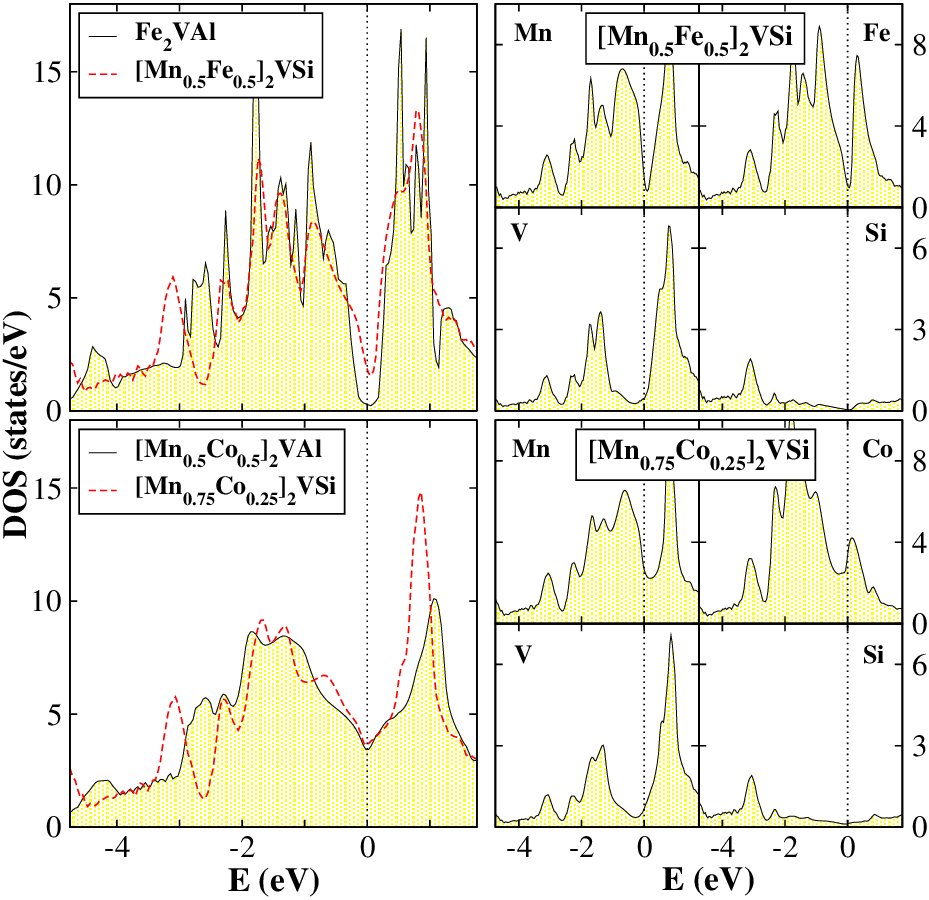}
\caption{(Color online) Left column : total DOS   for the
semi-metals Fe$_2$VAl and  [Mn$_{0.5}$Fe$_{0.5}$]$_2$VSi (upper
panel) and for non-magnetic calculations of the
[Mn$_{0.5}$Co$_{0.5}$]$_2$VAl and [Mn$_{0.75}$Co$_{0.25}$]$_2$VSi
alloys (bottom panel). All four compounds have
24 valence electrons in the unit cell.\\
Right column : atom-resolved DOS (scaled to one atom) for the
[Mn$_{0.5}$Fe$_{0.5}$]$_2$VSi  and [Mn$_{0.75}$Co$_{0.25}$]$_2$VSi
compounds. \label{fig3}}
\end{figure}

In the second part of our study we have investigated the effect of
using Fe instead of Co. In the bottom panel of Fig. \ref{fig1} we
include the DOS's for several concentrations and in Table
\ref{table2} the total and atomic spin moments. The conclusions
already drawn for the case of Co-doping are valid also for the
case of Fe doping. In the case of doping of Mn$_2$VAl the Fe
moment is parallel to the Mn one and very small ($\sim$0.2
$\mu_B$) while the case of [Mn$_{1-x}$Fe$_x$VSi] is similar to the
Co case with Fe moment antiparallel to the Mn one. As we increase
the concentration in Fe and reach Fe$_2$VAl and
[Mn$_{0.5}$Fe$_{0.5}$]$_2$VSi, which have 24 valence electrons,
the total spin moment vanishes. But our calculations indicate that
instead of a HMA we get a non magnetic compound.  To make the
origin of this different behavior clear we present in Fig.
\ref{fig3} the calculated DOS's for these compounds together with
non-magnetic calculations for the Co compounds. In Fe compounds
the Fermi level falls within a pseudogap and the alloys act as the
usual semi-metals (these results agree with previous calculations
by Weht and Pickett \cite{Weht2} while experiments suggest that
Fe$_2$VAl exhibits heavy-fermionic behavior being at the edge of
becoming magnetic \cite{Fe2VAl} but such a discussion exceeds the
scope of the present paper). Contrary, in the case of the
non-magnetic Co-compounds, the Fermi level falls within a region
of high DOS and due to the Stoner criterion the alloys prefer
energetically the magnetic configuration. In the right column of
Fig. \ref{fig3} we present also the atomic DOS. V atoms have the
same behavior in both cases and the high DOS for the Co-compounds
arises from the Co-Mn hybrids. In Ref. \onlinecite{Gala-Full} it
was shown that the gap arises between the occupied $t_{1u}$ and
the unoccupied $e_u$ states which are exclusively localized in
space at the higher valent transition metal atoms, here the Fe-Mn
or Co-Mn sites. In the case of Fe-compounds, these states are well
separated and the compound is a semi-metal. In the case of the
Co-compounds, if they were non-magnetic, these states strongly
overlap due to the different position of the Co-Mn hybrids
resulting in the high DOS at the Fermi level and the alloys prefer
the magnetic state (in Refs. \onlinecite{Galanakis-Review} and
\onlinecite{Gala-Full} it was thoroughly investigated why this
magnetic state prefers to be half-metallic).

We have studied the effect of doping the half-metallic
ferrimagnets Mn$_2$VAl and Mn$_2$VSi. Both Fe and Co substitution
for Mn keeps the half-metallic character of the parent compounds.
When the total number of valence electrons reaches the 24, the
total spin moment vanishes as predicted by the Slater-Pauling
rule. Whilst in the case of Fe-doping the
24-valence-electrons-compounds are non-magnetic semi-metals, in
the case of Co-doping half-metallic antiferromagnetism is
achieved. The driving force is the different position of the
states exclusively  composed by Mn-Co hybrids which strongly
overlap leading to very high values of the density of states at
the Fermi level for the non-magnetic phase and thus fulfilling the
Stoner criterion for the appearance of magnetism. Thus we have
presented an alternative way to create half-metallic
antiferromagnets for realistic spintronic application by simply
introducing Co atoms in the Mn$_2$VAl and Mn$_2$VSi half-metallic
ferrimagnets. Since crystals and films of both Mn$_2$VAl and
Co$_2$VAl alloys have been grown experimentally we expect these
results to stimulate a strong interest in both the theoretical and
experimental research in the emerging field of spintronics.



\begin{thebibliography}{999}

\bibitem{Zutic}
I. \v{Z}uti\'c, J. Fabian, and S. Das Sarma, Rev.
 Mod. Phys. \textbf{76}, 323 (2004).

\bibitem{deGroot}
R. A. de Groot. F. M. Mueller, P. G. van Engen, and K. H. J.
Buschow, Phys. Rev. Lett. \textbf{50}, 2024 (1983).

\bibitem{Galanakis-Review}
I. Galanakis,  Ph. Mavropoulos, and  P. H. Dederichs, J. Phys. D:
Appl. Phys. \textbf{39}, 765 (2006); I. Galanakis and Ph.
Mavropoulos, J. Phys.: Condens. Matter in press [preprint:
cond-mat/0610827].

\bibitem{Soulen}
R. J. Soulen Jr. \textit{et al.}, Science \textbf{282}, 85 (1998);
J.-H. Park, E. Vescovo, H.-J Kim, C. Kwon, R. Ramesh, and T.
Venkatesan, Nature \textbf{392}, 794 (1998).

\bibitem{McDonald}
T. Jungwirth, J. Sinova, J. Ma\v sek, J. Ku\v cera, and A. H.
MacDonald, Rev. Mod. Phys. \textbf{78}, 809 (2006).

\bibitem{Mavropoulos-Review}
Ph. Mavropoulos and I. Galanakis, J. Phys.: Condens. Matter in
press [preprint: cond-mat/0611006].

\bibitem{Nanda}
B. R. K. Nanda and I. Dasgupta, J. Phys.: Condens. Matter
\textbf{17}, 5037 (2005).

\bibitem{Leuken}
H. van Leuken and R. A. de Groot, Phys. Rev. Lett.  \textbf{74},
1171 (1995).

\bibitem{Felser2006}
S. Wurmehl, H. C. Kandpal, G. H. Fecher, and C. Felser, J. Phys.:
Condens. Matter \textbf{18}, 6171 (2006).

\bibitem{deGroot2}
R. A. de Groot, A. M. van der Kraan, and K. H. J. Buschow, J.
Magn. Magn. Mater. \textbf{61}, 330 (1986).

\bibitem{Kemal}
K. \"Ozdo\~gan, I. Galanakis, E. \c Sa\c s\i o\~glu, and B. Akta\c
s,  J. Phys.: Condens. Matter  \textbf{18}, 2905 (2006).

\bibitem{Gala-Full}
I. Galanakis, P. H. Dederichs, and N. Papanikolaou, Phys. Rev. B
\textbf{66}, 174429 (2002).

\bibitem{Sasioglu}
E. \c Sa\c s\i o\~glu, L. M. Sandratskii, and P. Bruno, J. Phys.:
Condens. Matter \textbf{17}, 995 (2005).

\bibitem{Akai}
H. Akai and M. Ogura, Phys. Rev. Lett. \textbf{97}, 026401 (2006).

\bibitem{Galanakis-RC}
I. Galanakis, K. \"Ozdo\~gan,  E. \c Sa\c s\i o\~glu, and B.
Akta\c s, Phys. Rev. B \textbf{74}, 140408(R) (2006).

\bibitem{itoh}
Y. Yoshida, M. Kawakami, and T. Nakamichi, J. Phys. Soc. Jpn.
\textbf{50}, 2203 (1981); H. Itoh, T. Nakamichi, Y. Yamaguchi and
N. Kazama, Trans. Jpn. Inst. Met. \textbf{24}, 265 (1983); C.
Jiang, M. Venkatesan, and J. M. D. Coey, Sol. St. Commun.
\textbf{118}, 513 (2001).

\bibitem{Weht}
S. Ishida,  S. Asano, and J. Ishida, J. Phys. Soc. Jpn.
\textbf{53}, 2718 (1984); R. Weht and  W. E. Pickett, Phys. Rev. B
\textbf{60}, 13006 (1999).

\bibitem{koepernik}
K. Koepernik and H. Eschrig, Phys. Rev. B \textbf{59}, 1743
(1999); K. Koepernik, B. Velicky, R. Hayn, and H. Eschrig, Phys.
Rev. B \textbf{58}, 6944 (1998).

\bibitem{Carbonari}
A. W. Carbonari, W. Pendl Jr., R. N. Attili, and R. N. Saxena,
Hyperf. Interactions \textbf{80}, 971 (1993).

\bibitem{Weht2}
R. Weht and  W. E. Pickett, Phys. Rev. B \textbf{58}, 6855 (1998).

\bibitem{Fe2VAl}
C. S. Lue, J. H. Ross Jr., C. F. Chang, and H. D. Yang, Phys. Rev.
B \textbf{60}, 13941 (1999).



\end{thebibliography}
\end{document}